# Spatial resolution(s) in atom probe tomography

Baptiste Gault[1,2], Frédéric De Geuser[3], Christoph Freysoldt[2], Benjamin Klaes[1], François Vurpillot[1]

[1] Univ Rouen Normandie, CNRS, INSA Rouen Normandie, Groupe de Physique des Matériaux, UMR 6634, F-76000 Rouen, France
[2] Max-Planck-Institute for Sustainable Materials, Max-Planck-Str. 1, Düsseldorf 40237, Germany
[3] University of Grenoble Alpes, CNRS, Grenoble INP, SIMaP, Grenoble, F-38000, France

## Abstract

Atom probe tomography (APT) is often quoted to provide "atomic-scale" analysis of materials in three dimensions. Despite efforts to quantify APT's spatial resolution, misunderstanding remain regarding its true spatial performance. If the depth resolution was once reported to be 20 pm, quoting this value outside of its specific context is misleading and should be avoided. The resolution achievable in pure metals, at one specific location within one reconstructed dataset, does not generally apply across materials or analysis conditions, or even throughout a single tomographic reconstruction. Here, we review various efforts at defining and measuring the spatial resolution in the study of single phase and single element materials – i.e. pure metals – in field-ion microscopy (FIM) and APT. We also report on the degradation of the resolution arising from ion optical devices used to improve the mass-resolution. We aim to offer some perspective as to how reported resolutions may be or may not be of any relevance to most of the materials characterisation efforts by APT, including cases of precipitates in a matrix that emphasise the need to consider an effective resolution. Finally, we discuss concepts to improve the spatial accuracy of the technique in a relatively distant future.

# Introduction

Over the past couple of decades, atom probe tomography (APT) has been rising in prominence within the cornucopia of materials characterization techniques (Devaraj et al., 2018; Kelly & Miller, 2007; Gault et al., 2021; Lefebvre-Ulrikson et al., 2016). Although often presented as a microscopy technique more so than a microanalysis technique, we would argue that APT is both and neither in the most traditional sense. One could even argue that APT is a lousy mass spectrometer, with mass resolving power inferior to most other techniques because of the relatively shorts times-of-flight imposed by the need to maintain a sufficiently large collection angle (Kelly et al., 2004; Deconihout et al., 2007). The relatively low number of counts – in the millions to low billions – also limits the sensitivity (Haley et al., 2020). APT is also a lousy microscope in which the specimen itself is the main lens (Müller, 1970) and that is plagued by aberrations that cannot be corrected as they arise from the physics of field evaporation itself. The true strength of APT is in the combination of microscopy and microanalysis enabling compositional mapping in 3D on an unparalleled scale (Blavette et al., 1993).

Conversely to what is sometimes claimed, APT does not allow for "atomic-scale analysis" that would imply somehow that the spatial resolution allows for resolving the true position of individual atoms. Such ideas may arise from considerations from the precursor technique of APT, field-ion microscopy (FIM) in which atoms are effectively imaged on some but not all of the atomic terraces associated to sets of atomic planes (Müller & Bahadur, 1956; Chen & Seidman, 1971). This may also be caused by a lack of an "absolute" definition of the spatial resolution of APT – and the misconception that such a value could even exist, as will be discussed in the following.

Multiple definitions and approaches to measure the resolutions have been proposed (Vurpillot et al., 2001; Kelly et al., 2007; Gault et al., 2009) based on experimental data for pure materials, either Al, W, Fe, or doped Si (Cadel et al., 2009), for which the field evaporation tends to be relatively well-behaved. Non-surprisingly maybe, they fail to agree on the specific ways to quantify the resolution, yet they conceptually broadly agree on the fact that the spatial resolution is not isotropic, varies across the field of view and fundamentally limited by irregularities in the sequence of field evaporation and trajectory aberrations. This is not new: it has been known for five decades at least (Krishnaswamy et al., 1975; Waugh, A R et al., 1976), it is supported also by extensive simulation work (Vurpillot, Bostel, & Blavette, 2000; Vurpillot & Oberdorfer, 2015; Ashton et al., 2020).

The case of non pure-metals is more complex as additional aberrations can occur both laterally and in the depth. Recent experimental work by analytical FIM allowed for quantifying aberrations of up to over 0.5 nm for isolated W atoms in Ni, so the distance to the 4th nearest neighbour in the Ni-matrix (F. Morgado et al., 2025). This agrees with simulations performed of a similar system, i.e. a Ni-Re binary, with the model introduced by (Nicolas Rolland et al., 2015) that reported that for neighbouring Re atoms in the initial synthetic sample, i.e. with a separation of less than 0.2nm, after field evaporation and reconstruction, the distance separating these two atoms was typically near 0.5 nm (Gault et al., 2022). The case of more compositionally-complex materials is also discussed therein based on similar simulations. Limited reports on ceramics for instance reported a lateral resolution of 1.5 nm in SiC (Ndiaye et al., 2023). As discussed for multilayers for instance (Vurpillot et al., 2004; Marquis et al., 2011)  or for precipitates in a matrix (Vurpillot, Bostel, & Blavette, 2000; De Geuser & Gault, 2020) additional complexities arise from mesoscale changes of the curvature.

However, in most of these cases, quoting the depth or even lateral resolution measured for a pure metal simply makes no sense. And despite a consensus that APT does not have "atomic resolution" and cannot resolve interatomic distances in three-dimensions – the lateral resolution is simply too low – the literature is still plagued with reports of studies based on interatomic distances that are, if not erroneous, at the very least misleading, probably unphysical and setting unrealistic expectations for non-experts.

This motivated us to get back to provide a perspective on this body of work on the spatial resolution and the resolution limits across instruments and experimental parameters. We argue that no single value of the spatial resolution in APT can be quoted, even within a single dataset, as across the field-of-view, there is a range of values. This range varies even in the analysis of a single specimen, more so across specimens, and across materials, and reducing this complexity to a single value simply leads to misconceptions and should be avoided. Finally, as a community, we should fight the idea that APT can generally resolve atomic neighbourhood relationships, and only in specific cases can this be achieved. That does not remove anything from the uniqueness or importance of APT in the cornucopia of materials characterisation techniques, simply that it needs to be considered for its strengths and with its own intrinsic limitations.

# Field ion microscopy

It seemed important to start this perspective with FIM, from where we feel some of the misconceptions on APT's resolution might be arising. In FIM, the image of a single atom is made of thousands of gas ions striking the screen per second. Under ideal ionization conditions of the imaging gas, the resolution can be assumed to be the diameter of a single bright spot produced by a single atom on an image, and hence depends on the specimen's radius of curvature (r_0), the magnification and hence the shape of the specimen and resulting ion compression (β), as well as the velocity of the gas ions (ϵ_T) and hence the specimen's base temperature and that of the imaging gas as well as the nature of the gas used (H_2,He,Ne,Ar,…). An expression for the resolution $δ_S$ was proposed by (de Castilho & Kingham, 1987), based on an earlier formulation proposed by (Chen & Seidman, 1971):

$$\delta_s = \left( \delta_0^2 + \frac{2\beta h r_0 k_f^{\frac{1}{2}}}{\pi(2meV_0)} + 16 k_f \left( \frac{\beta^2 r_0^2}{eV_0} \right) \varepsilon_T \right)^{1/2}$$

where $δ_0$ is the diameter of the ionization zone above a given atom, *m* the mass of the imaging gas ion, $k_f$ the geometrical field factor, β is the image compression factor, $ε_T$ the transverse thermal energy associated to the image atom gas when it is ionized, *h* the Planck constant and $V_0$ the electrostatic potential.

Since the image resolution depends on these many parameters, including radius, temperature, gas used, quoting a single value is already impossible. In the case of helium at typical cryogenic temperatures, the resolution is on the order of 0.20 nm to 0.25 nm (Miller et al., 1996). The best results are obtained for specimens with small radii at low temperatures (10 K–50 K), using helium or neon as imaging gases. So, this resolution represents the minimum distance below which two adjacent bright spots created by two different atoms can no longer be distinguished as separate. This also means that at a constant $\delta_S$ across the field-of-view, the packing of the atoms on certain planes will preclude imaging of individual atoms, Figure 1.

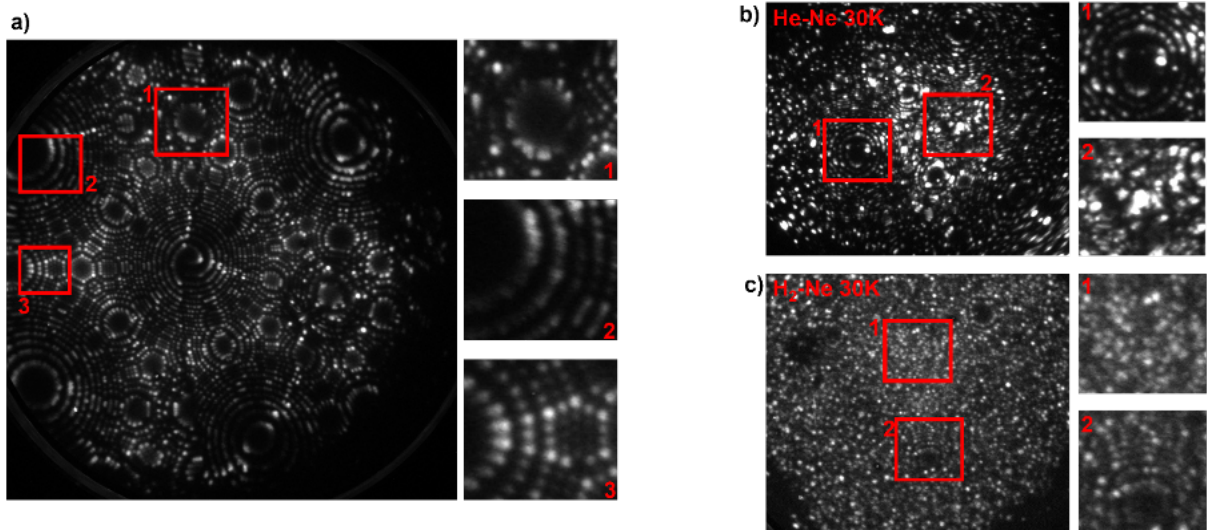

*Figure 1: (a) W 50K, pulse 45%, He100%. (b) ODS 30K, He75%-Ne25%. (c) ODS, 30K, H225%-Ne75%.*

Beyond this theoretical limit, there are additional constraints arising from the instrument itself. Indeed, it is also necessary to consider the precision with which the position of a spot can be localized within the image. The position of an atom is approximated by the centre of the spot it produces. The accuracy of determining this centre depends on the image pixel density and the method used to compute the position. Typically, this localization precision is around ±0.05 nm for a 1024 × 1024 pixels image. When the pixel intensities within the spot are used to weight each pixel's contribution in the centroid calculation, the localization accuracy is slightly improved (Miller et al., 1996).

These are important considerations when developing algorithms to process FIM images, particularly for 3D FIM as well as for identifying structural and microstructural defects in images and moving towards analytical FIM. The image resolution and the accuracy of atomic position determination will therefore affect the lateral resolution in 3D FIM reconstructions, which are obtained from sequences of surface images acquired regularly during the specimen's evaporation, as described in (Klaes et al., 2021). Therein they determined the best achievable resolutions and most accurate reconstruction for pure tungsten, with resolutions calculated by the method based on Fourier Transform introduced by Vurpillot et al. (Vurpillot et al., 2001) and modified in (Lefebvre-Ulrikson et al., 2016) as discussed below. During these analyses, the specimen is imaged using ionisation at the DC voltage, and the field evaporation is triggered by nanosecond voltage pulses superimposed on this DC voltage. At 50 K, with a pulse fraction of 45%, depth resolution can be approximately 0.05 nm and lateral resolution down to 0.1 nm (Figure 2a-b). These results are consistent with the evolution of the calculated resolution as a function of temperature reported in (Miller et al., 1996), and demonstrate that, under optimal analysis conditions, it is possible to achieve true three-dimensional atomic resolution on 011 planes (Figure 2c).

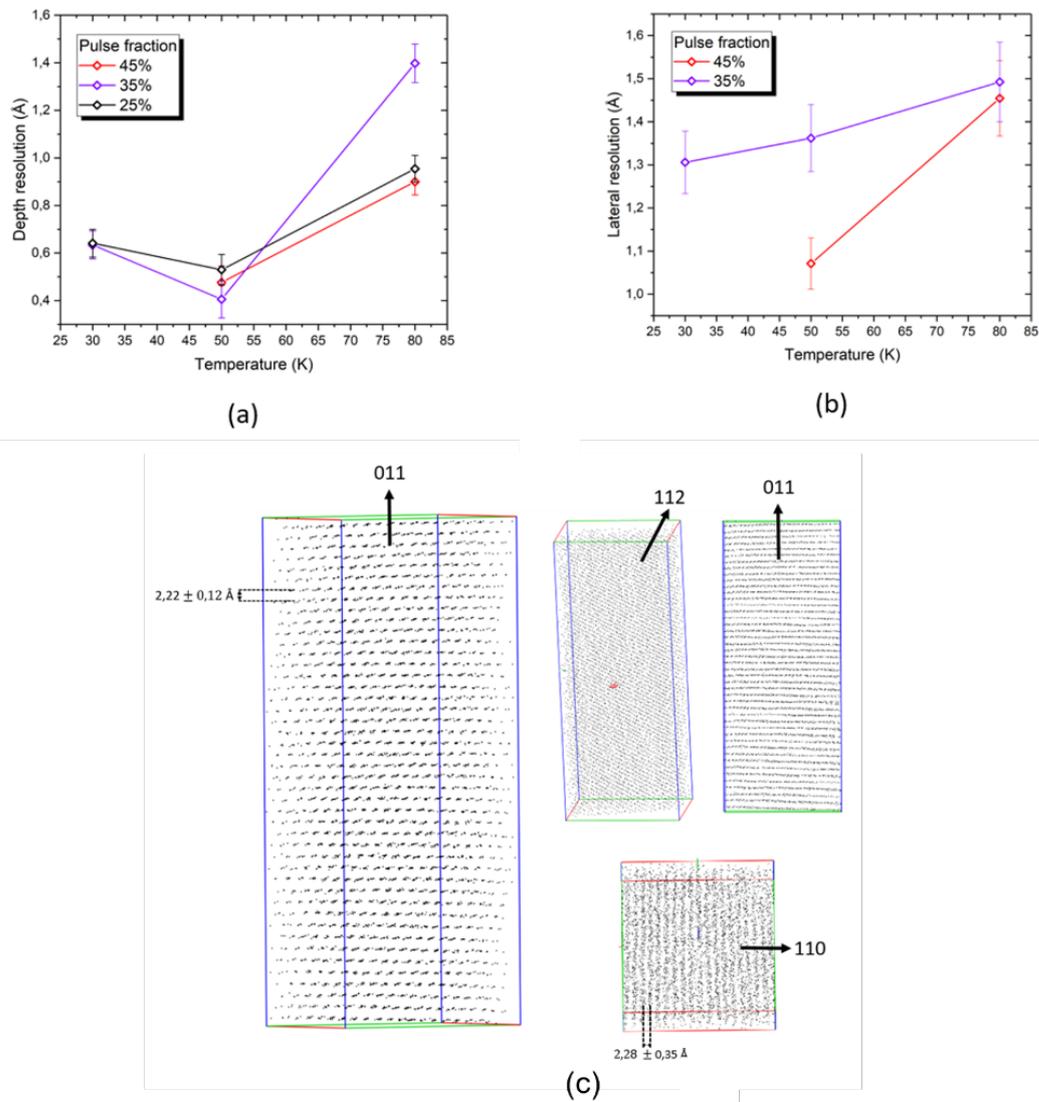

*Figure 2: (a) Depth resolution on W in function of T°. (b) Lateral resolution on W as a function of the base temperature. (c) Small volume of W reconstruction analyses at 50K with true 3D atomic resolution (from (Klaes et al., 2021)).*

In this study, the method used to calculate the depth increment associated with each image is that proposed by Dagan et al. (Dagan et al., 2017). This approach is based on counting the number of images required to completely evaporate a surface atomic layer along a given crystallographic direction (here, <011>). This method proves to be highly accurate when one of the poles is well resolved and clearly distinguishable in the images, as it relies on the real reticular distance. However, for materials in which the poles are not sufficiently visible in images, or when the relevant parameters are unknown, using the procedure proposed by Bas et al. (Bas et al., 1995) to determine the depth increment may introduce additional uncertainty in the atomic position along the analysis direction, thereby degrading the resolution – particularly in depth.

However, this case study on a pure metal can serve as a reference to determine the optimal resolution values that can be achieved and to establish, as in APT, a lower limit. The complexity of this material remains moderate, as it contains only two chemical species with relatively similar evaporation fields (~33 to 35 V/nm). Other 3D FIM analyses on oxide-dispersion (Fig1.b-c) strengthened ferritic or ferritic-martensitic steels containing a uniform dispersion of nano-sized oxide particles, typically $YTiO_2$ or $Y_2TiO_5$ for this material (Hatzoglou et al., 2017), which act as strong obstacles to dislocation motion, have reported for the ferritic matrix a depth resolution on the order of 0.05 nm and a lateral resolution around 1.5 nm. Phenomena such as surface atom migration shown in SiC (Ndiaye et al., 2023), contrast and spot-size variations between different chemical species (Klaes et al., 2022), or surface recombination (Veret et al., 2025) can all affect and degrade the resolution.

Although the image resolution generally allows individual atoms to be distinguished and separated in most cases, several factors limit the overall resolution of 3D reconstructions in FIM. These include the accuracy of atomic position determination on the image — which depends, among other things, on the number and size of pixels and on the sharpness of the spot associated with each atomic position — as well as surface atom mobility and surface molecular recombination.

Finally, approaches have been proposed that combine FIM for its positioning accuracy and APT for elemental identification (Katnagallu et al., 2019, 2022; Klaes et al., 2022). Analytical FIM has not yet managed to deliver routine atomically-resolved analytical mapping though, and much work remains to be done to demonstrate that this approach can be generalised beyond selected binary alloys(Morgado et al., 2021; F. Morgado et al., 2025), as the high level of background can preclude identification of the field evaporated ions. First results still provide means to discuss the amplitude of aberrations in pure metals (Gault et al., 2022) and binary alloys(F. Morgado et al., 2025), that demonstrate more clearly that the lateral resolution after field evaporation is insufficient to maintain first nearest neighbour relationships on average.

# Definitions of the spatial resolution

Let us now shift the focus from FIM to APT. Vurpillot et al. introduced a definition of the spatial resolution in the depth as the width of a set of reconstructed atom planes from a selected family (Vurpillot et al., 2001), which could be readily calculated from the direct Fourier transform. Assuming the spatial resolution is a sum of a lateral resolution, i.e. tangential to the surface of the specimen, and an in-depth term, they proposed that the lateral resolution can be determined by probing the thickening of the imaged atomic planes this set of planes at a known angle. Indeed, the thickness will change because of a contribution from the lateral offset of the atoms in the reconstruction compared to their expected position (Figure 2). Note that since both lateral and depth dispersions are physically independent, it was further assumed that the final plane resolution is a quadratic sum of both terms (Lefebvre-Ulrikson et al., 2016).

Figure 3 details the measurement of the spatial resolution within a pure Al specimen analysed in a Cameca LEAP5000XS, in VP mode, 15% PF, 40K, 2% DR. A subset of $10^6$ atoms is displayed in Figure 3 a, and a region-of-interest (ROI)indicated by the black rectangle is extracted. Figure 3 b show the set of (113) atomic planes within the ROI. The theoretical impact density through this set of atomic planes is a Dirac comb convoluted by a point spread function (PSF) with a half maximum width Δ, as schematised in Figure 3 c. the direct Fourier transform of b) is a Fourier comb in the reciprocal space multiplied by the Fourier transform of the PSF, plotted in Figure 3 d. An equivalent of the spatial distribution map (SDM) (Geiser et al., 2007; Moody et al., 2009) can be calculated as the inverse Fourier transform of the intensity of the Fourier transform in Figure 3 d, in other words the autocorrelation function. Measuring the amplitude of the 1$^{st}$ diffraction peak (at k=1/d$_{hkl}$) enables measurement of the width of the PSF, and thus the local plane resolution, using the Gaussian approximation such that:

$$\Delta = 2\sqrt{2ln2}\sigma\ 2.35 \left( \frac{d_{hkl}\sqrt{2ln\left(\frac{1}{Fhkl}\right)}}{2\pi} \right)$$

where σ is the standard deviation of the Gaussian function. The resulting measurements of the variations of the planes thickness as a function of θ, the angle between the set of planes and the normal to the specimen's surface, $\Delta(\theta)$, is plotted in Figure 3 e. The experimental data is fitted assuming a quadratic sum of a lateral resolution ΔL of 500±100pm and in-depth resolution Δ of 35±5 pm.

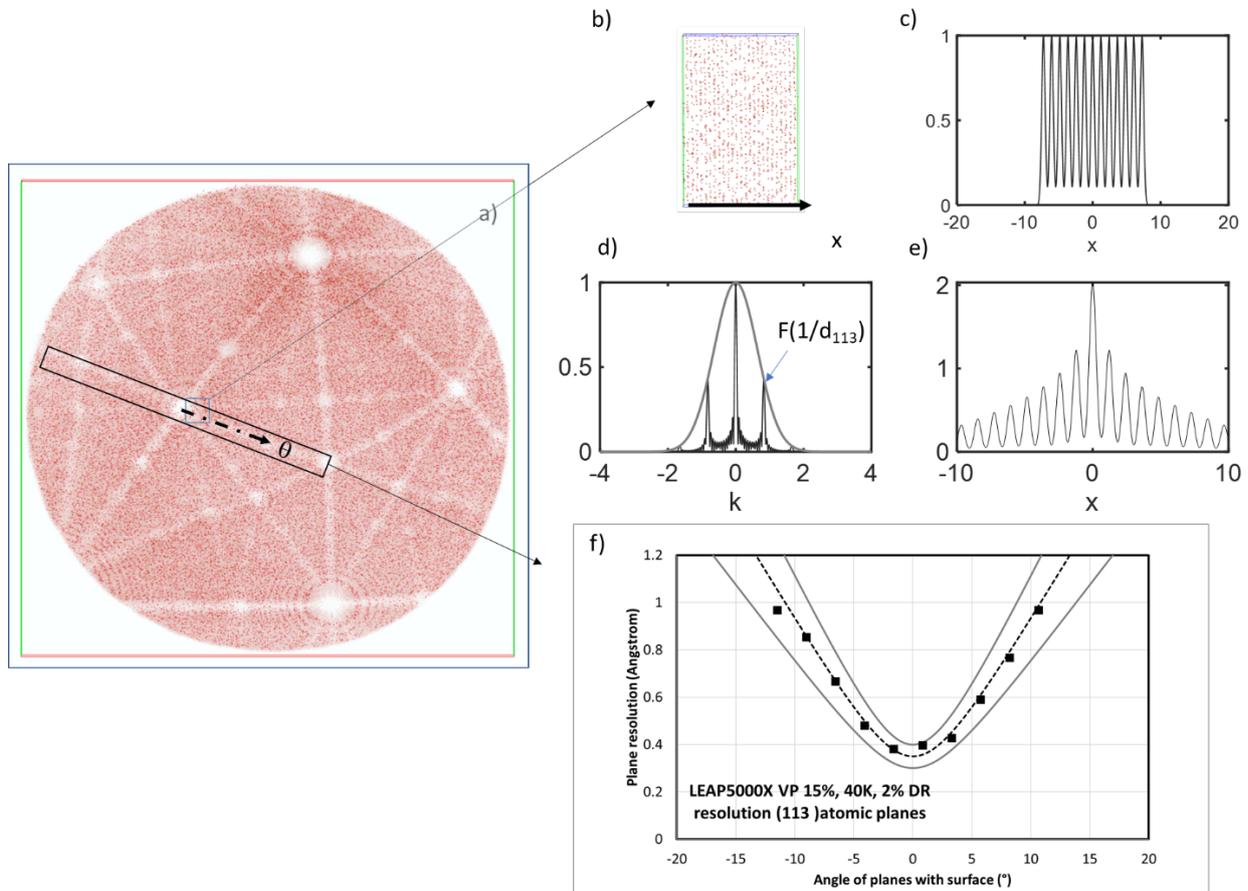

*Figure 3: a) (x,y) slice through a pure Al dataset containing $10^6$ ions. b) set of (113) atomic planes within the region-of-interest marked by the black box in a). c) Theoretical point density*

*through a set of atomic planes. d) Direct Fourier transform of b) and e) corresponding spatial distribution map. f) Experimental measurements of Δ(θ) fitted a quadratic sum of a lateral ΔL and in-depth Δp resolutions of 500±100pm and 35±5 pm, respectively.*

In pure metals, the depth resolution measured by this method was found in the sub-angstrom range, and the lateral resolution was found to approach the first neighbour interatomic distance, as demonstrated for pure tungsten in Figure 4. The top-view on the reconstructed dataset in Figure 4a contains a ROI shown as a red square that is displayed in Figure 4b. The corresponding Fourier transform in Figure 4c shows the equivalent of diffraction spots corresponding to multiple sets of atomic planes imaged within this ROI with a lateral resolution in the range of 0.2 nm.

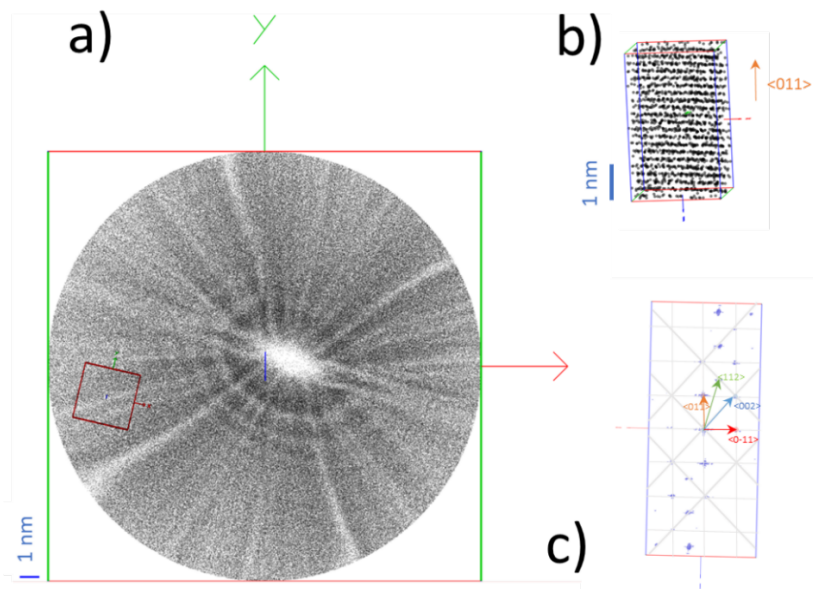

*Figure 4: a) pure Tungsten analysis in APT (LEAP 3000 analysis in voltage pulsed mode at 58K) b) Small subset of the volume oriented along the <011> crystallographic direction (close to the specimen z axis) c) Projection of the 3D Fourier transform of the ROI highlighting with the equivalent of low Miller indices diffraction spots in the (x,y) plane showcasing a lateral resolution in the range of 0.2 nm.*

For the depth resolution, a similar definition was used by Gault et al. (Gault et al., 2009) but extracted from SDMs. At this stage it is likely important to recall that the way an SDM is typically calculated (Boll et al., 2007; Geiser et al., 2007) is exactly equivalent to pair or radial distribution functions (Marquis, 2002; Sudbrack et al., 2006; De Geuser et al., 2006), simply it is split to provide a view of the interatomic distances along the reconstructed depth (z-SDM) and in two-dimensions within these planes (xy-SDM), and only for the purpose of visualisation. The z-SDM is hence related to the radial distribution and the Fourier transform of a given set of planes as pointed in Figure 3d–e.

To measure the resolution, the central peak of the z-SDM obtained in a ROI centred on one of the observed poles is fitted with a Gaussian function following a background reduction. The background reduction makes relatively little difference to the derived depth resolution (Moody et al., 2009). The lateral resolution was quantified from the xy-SDM calculated across the entire field of view, also following background reduction, but in this case, as shown in, the signal-to-background ratio is low and most of the atoms – an estimate would suggest at least 99% for this (200) set of planes – end up not reconstructed with the expected interatomic distance. We may note that the xy-SDM spot positions are the result of the multiplication of different lattice plane signals coming from different locations over the tip surface. Based on a Fourier transform of the SDM, Kelly et al. (Kelly et al., 2007) introduced yet another definition of resolution, based on the measurement in the reciprocal space of the position of the farthest diffraction spot which is more akin to a resolution limit, and as such slightly different to the other two definitions.

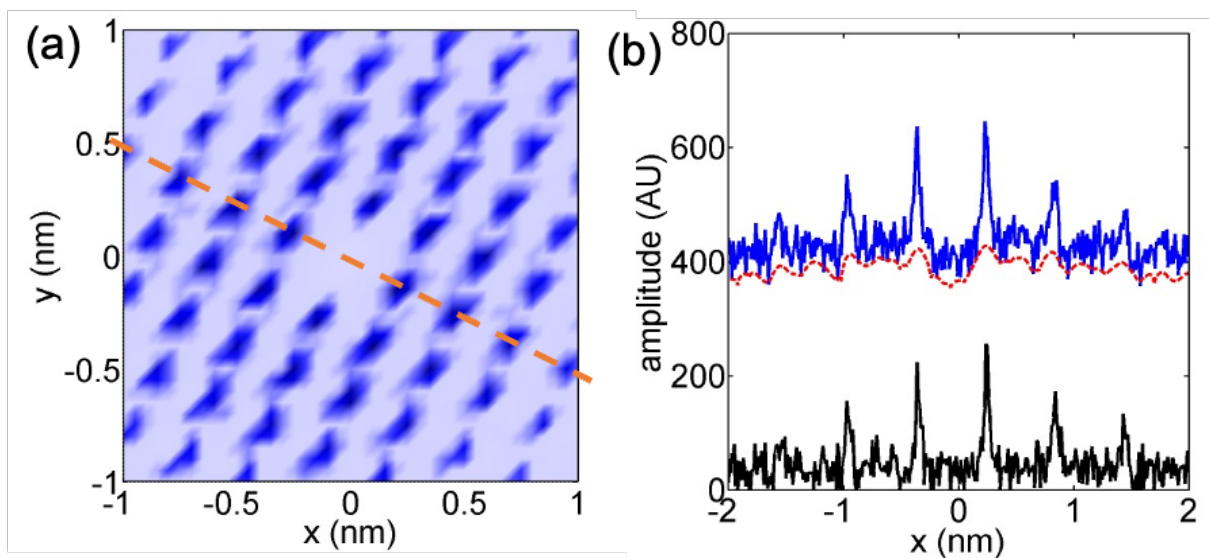

*Figure 5: (a) 2D spatial distribution map for {002} planes in pure Al analysed at 40K; and (b) line profile of the amplitude along the orange dashed line plotted in blue. The background estimation is in red, and in black is the background subtracted signal that is used for estimating the lateral resolution (figure modified from (Moody et al., 2009)).*

In microelectronics, the spatial resolution is often quantified through the decay length measured in nm/dec, which is the distance over which the measured concentration decreases by one decade (i.e., by a factor of 10) when crossing an interface and commonly used in depth profiling in secondary-ion mass spectrometry (SIMS) (Shimizu et al., 2009). In other words, the unit nm/dec therefore expresses how many nanometres it takes for the signal to fall by one order of magnitude. This is measured experimentally by extracting a profile across the interface, and plots concentration on a log scale, and fits a line in the exponential tail region. If decay length is induced by a convolution by a gaussian spatial spread, the decay length is mathematically close to the $\sigma$ value. In Si/SiGe interfaces, Koelling et al. measured decay length in the in-depth direction as good as 0.2 nm/decade (Koelling et al., 2009), within the range of reported depth resolutions including for Si (Cadel et al., 2009). They found that the decay length increases when higher laser pulse energies and longer laser wavelengths are used due to enhanced preferential evaporation and the development of a non-hemispherical specimen end shapes leading to additional aberrations. Similar measurements were performed by Shimizu et al. in $Si^{28}/Si^{30}$ multilayers heterostructures in laser-assisted atom probe and a direct comparison between the depth and lateral directions was performed (Shimizu et al., 2011). The best average decay length in depth was found to reach 0.3 nm/decade whereas the lateral is degraded to 1.3 nm/decade.

Finally, De Geuser et al. proposed to consider an effective spatial resolution suitable for the analysis of small particles (De Geuser & Gault, 2020). The radial distribution functions calculated from the reconstruction of datasets containing small precipitates were processed by using a formalism similar to small-angle scattering (Zhao et al., 2018). In such a case, the radial distribution functions are not used to probe the first shells of nearest neighbours, but to quantify the composition field up to ten or more nanometres around each atom. The direct correlation with small-angle scattering on the same material showcased a cut-off of about 1 nm, below which the particles' size can no longer be estimated. Since the interface of small particles average the spatial resolution in all directions, a single effective spatial resolution was introduced ($\sigma_{eff}$). The effective spatial resolution should be considered as a rotational average of the resolution in all directions. In this case, the average effective resolution can be expressed as:

$$\sigma_{eff} = \sqrt{\frac{2\sigma^2_L + \sigma^2_p}{3}}$$

Where $\sigma_L$ and $\sigma_p$ the lateral standard deviations of the lateral resolution and in-depth resolution (note that the fullwidth half maximum of each resolution are given by $\Delta \sim 2.35\sigma$ assuming a normal distribution of positioning uncertainty).

## So is APT's spatial resolution 20 pm?

Gault et al. reported that it could be 20 pm in depth for the (206) planes (Gault, Moody, et al., 2010). That value though is derived in that specific ROI within that dataset. For this same set of planes in this same section of the reconstructed data, the lateral resolution is 200 pm, an order of magnitude worse. Claiming that the spatial resolution is 20 pm is hence misleading. Can these values be reached again? Yes, most likely in similar experimental conditions (pure Al, 40K, shank angle of ca. 5°, radius of ca. 60 nm), but already in the same article, there is a demonstration that for another specimen, the same set of planes exhibits a different resolution. Can these values be used outside of this specific case? Strictly speaking, no. It does not even apply 5 or 10 nm away to the side of this region of interest within the same dataset.

Follows the question of what to quote when asked "what is the spatial resolution of APT?". There is sadly no easy answer to this question. And ultimately, the better question would be "what are the spatial resolutions of APT?". Even considering the idea of a resolution limit can be misleading – 20 pm may never be reached on a specific material because of its compositional complexity. Or, in the case of an alloy containing particles, it may be reached in specific regions of the matrix but not for the particles. And APT is typically used to measure the composition of these particles.

# There is not one resolution but resolutions in APT

In summary, the aforementioned body of work demonstrated that:

- The resolution both in depth and laterally is primarily associated to the local crystallography of the specimen. It can be measured in a region-of-interest containing a selected set of atomic planes, localised within the reconstructed point cloud. The selection can be facilitated by using the detector hit map or a map of the point-density in the reconstructed data (Vurpillot et al., 2001, 2003; Gault, Moody, et al., 2010);
- It is not best in the centre of the detector as sometimes assumed in the past because maybe of early work in FIM (de Castilho & Kingham, 1987), but where atomic planes are locally reconstructed (Gault, Moody, et al., 2010);
- The depth resolution is primarily controlled by the deterministic sequence in which atoms are field evaporated from the edges of the terraces at poles. The depth resolution is better (i.e. understand with a lower value) for sets of atomic planes with higher Miller indices, and this was assumed to be related to the relative size of the atomic terraces formed by the corresponding set of planes intersecting the curved emitting surface (Gault, Moody, et al., 2010). The depth resolution worsens relatively faster for smaller poles that correspond to sets of planes with a lower interspacing (Drechsler & Wolf, 1958);
- The lateral resolution is limited due to aberrations in the early stages of the flight caused by the distribution of the electrostatic potential in the vicinity of the edges of atomic terraces (Vurpillot et al., 1999; Oberdorfer et al., 2013; Nicolas Rolland et al., 2015), as well as by the roll-up motion of surface atoms on their neighbours prior to departing from the surface (Waugh et al., 1975; Suchorski et al., 1996; Ashton et al., 2020);
- The spatial resolution degrades with increasing temperatures, with the depth resolution relatively more robust; at higher temperature, surface diffusion-related processes get activated (Gault, Müller, et al., 2010; Vurpillot, Gruber, et al., 2009; Parviainen et al., 2016), leading to an increase to the background which is not considered in the report of the resolution;
- In laser pulsing mode, these processes are also activated, as the field evaporation takes places at relatively higher temperatures (Vurpillot, Houard, et al., 2009), and hence the spatial resolution is relatively worse. Recent measurements made in metals (Jenkins et al., 2024) or in non-metals (Ndiaye et al., 2023), showed a significant worsening of lateral resolution of up to 1–2 nm in pulsed laser APT.
- The presence of a reflectron or any device containing meshes on the path of the ions introduces an intrinsic blurring that degrades the lateral resolution. This blurring does not affect the depth resolution. Recent measurements in metals degradation of the lateral resolution up to a factor 2 (Jenkins et al., 2024), with a dependence on the size of the probed area.
- Compositional complexity makes this process less well determined causing degradation of the resolution (Vurpillot, Bostel, Cadel, et al., 2000; Gault et al., 2022). The resulting increased roughness of the specimen surface causes additional aberrations that contribute to deteriorating the lateral resolution (Gault et al., 2022). We may note that the resolution loss removes or reduces the presence of density variations induced by zone lines. This effect is worsened in non-metallic materials,

- probably because of the reorganisation of the surface during the field evaporation process (Ndiaye et al., 2023; Veret et al., 2025).
- If different chemical species require vastly different field strength for evaporation, the resolution may depend on the species. Atoms that are more difficult to evaporate stay longer on the surface in exposed configurations, enhancing the chance for vertical retention (Martin et al. 2019), lateral diffusion (Ohnuma, 2019), artificial clustering (Op de Beeck et al. 2021), or repeated drag towards steps until they accumulate at zone lines and poles (Kobayashi et al. 2011; Gault et al, 2012; Oberdorfer et al., 2018; Martin et al. 2019).

These are for relatively simple, single-phase materials. The presence of multiple phases, interfaces or layers of different composition lead to the formation of local curvatures that change the magnification and create trajectory crossings and cause overlap in the projected images of these different microstructural features onto the detector that result in overlaps of the compositional fields within the reconstructed data (Vurpillot, Bostel, & Blavette, 2000; Larson et al., 2013). For such cases, there is no current direct way to measure the resolution "that matters", that is the precision with which APT can image particles for instance, how distorted they might appear but also how precisely their composition can be measured. De Geuser et al. proposed to use a correlation with small-angle scattering techniques to derive what they referred to as an effective resolution that is typically in the range of 1 nm, depending on the composition of the particles and the matrix (De Geuser & Gault, 2020).

# Why does this matter to me as an APT user?

Well, measurements are made to obtain data. That data has an inherent precision and accuracy, based upon which one can derive an interpretation to turn this data into a result that can be integrated into a study. Not understanding the limits of a measurement can lead to misinterpretation. The idea is not to emphasise APT's limitations, but to trigger a reflection on what it can truly achieved – and this is a critical aspect of any metrology. APT does not provide true atomic resolution, and this means that some conclusions cannot be drawn.

## For interfacial segregation

Let us take an example showcasing that also those who claim to have the upper hand today have also made mistakes in the past that got published in the open, peer-review literature. In 2011, Gault et al. reported a study of an Al–Cu–Li–Mg–Ag alloy by APT, and one of the highlights of the paper was "Mg and Ag are partitioned inside the T1 phase but do not segregate at the interface of the plate-shaped precipitate with the matrix" (Gault et al., 2011). The precipitates of interest here are T1, one of the strengthening phase, that are plates with {111} habit planes. Specimens were prepared by electrochemical polishing from blanks cut from a plate of the material, and the orientation of the specimens was hence not controlled. All analyses performed on this set of samples were done without these planes in the field-of-view of the atom probe. A few years later, an alloy from the same series, with a slightly different composition, was studied in depth by Araullo-Peters et al. (Araullo-Peters et al., 2014). They report "Ag and Mg segregation to T1 interfaces was systematically observed when the plates were oriented perpendicular to the probing direction".

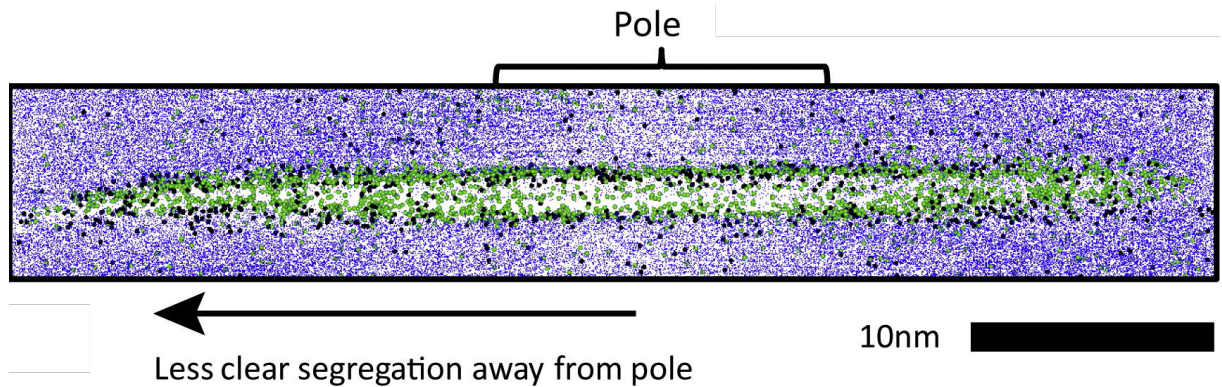

*Figure 6: Atom probe data cross section of a T1 plates found to lie on {1 1 1} planes in the sample after 84 h of aging. Al atoms are represented as blue dots; Mg and Ag are respectively green and black sphere (Figure from (Araullo-Peters et al., 2014)).*

In **Error! Reference source not found.**, the segregation of Mg and Ag at the interface between the matrix and the T1 precipitate is readily visible where the resolution is the highest, in the middle of the pole, and there is a "blurring" of the interfacial segregation away from the pole. This intermixing arises from the drop of the resolution making it nearly impossible to image and quantify the segregation. This effect led to the misinterpretation discussed above.

The lack of precision has also been fuelling many discussions to reconcile the compositional width measured by APT, that can be up to several nanometres on either side of a grain boundary, in comparison to its structural width boundaries or interphase interfaces measured by transmission-electron microscopy to be generally in the range of 0.5 – 1 nm or so. Revisiting data from (Danoix et al., 2016), Figure 7 a, that includes an interface between two ferrite grains with an orientation relationship making the two (011) poles almost superimposed, Figure 7b, Jenkins and co-worker could demonstrate segregation at an interface over only 4–5 atomic layers (Jenkins et al., 2020), Figure 7c. This width is compatible with other observations resulting in part from comparison with density-functional theory simulations (Varanasi et al., 2025). However, such a high resolution could be achieved because on either side of this interface is a (011) pole in the body-centered cubic matrix, along which the depth resolution is high. This may not be possible in a general case, but it makes sense to exploit the higher resolution in the depth whenever possible. Modern specimen preparation techniques based on focused-ion beam can be performed following preliminary characterization by e.g. electron backscatter diffraction for instance, and transmission Kikuchi diffraction to try and position the interface within the specimen to exploit the optimal resolution.

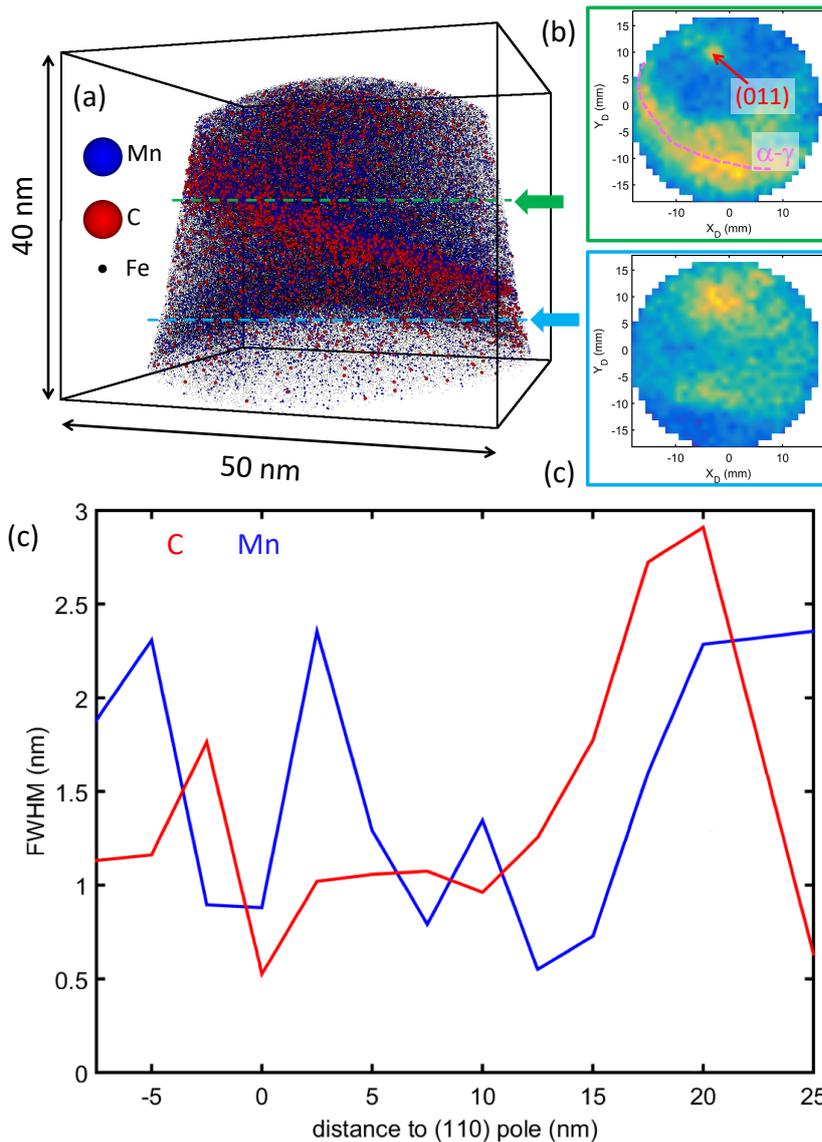

*Figure 7: Reconstructed APT map showing the distribution of Mn, C, and Fe in the dataset containing the interface. For clarity, only 5% of the Fe ions are displayed. (b,c) Detector hit maps calculated for a slice of 0.5 million ions at different depths indicated by the arrow of the corresponding color in (a). In (b), a pole is indicated with a red arrow and the position of the α–γ interface is marked by a pink dashed line. Width of the composition profile for C (red) and Mn (blue) as a function of the distance to the center of the (110) pole (from (Jenkins et al., 2020)).*

## Clustering and short-range ordering

For instance, data processing techniques that explore atomic neighborhoods in 3D typically do not consider variations in spatial resolution across the field-of-view – this includes $k^{th}$ nearest neighbour calculations (Shariq et al., 2007; Stephenson et al., 2007) and derivative cluster-finding algorithms (Marquis & Hyde, 2010; Dumitraschkewitz et al., 2018), as well as radial distribution functions (Sudbrack et al., 2006; De Geuser et al., 2006; Haley et al., 2009).

By using radial distribution functions, Haley and co-workers evidenced that some limited regions within the data, away from the poles, can retain some information on atomic neighbourhood relationships (Haley et al., 2009), albeit with distortions, as summarised in Figure 8. This result was only achieved in pure Al, and not demonstrated in other materials. As such, using such techniques for studying atomic neighbourhoods and discuss for instance short-range ordering, can be erroneous.

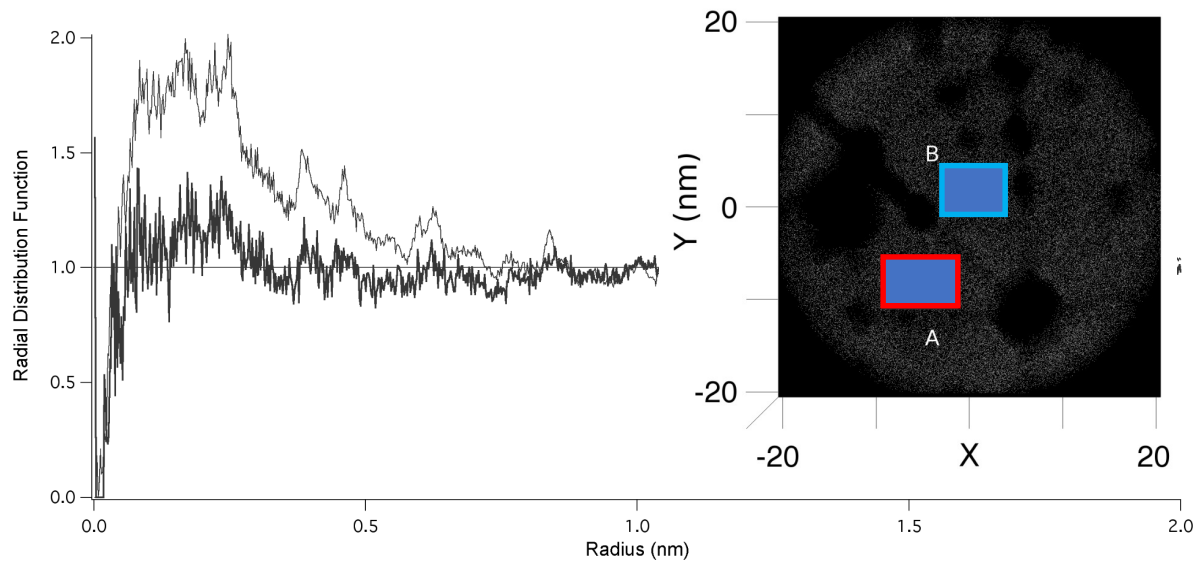

*Figure 8: RDF obtained from the pure aluminium sample from two regions (A & B) selected for analysis, each well away from any major zone line or pole (from (Haley et al., 2009)).*

For studying short-range ordering, this is somehow more complex, as this involves truly analysing interatomic distances in non-pure metals. There have been recent reports in compositionally complex alloys, with one approach making use of nearest neighbor distributions (He et al., 2024), as inspired by preliminary work from this group (Ceguerra et al., 2010; Moody et al., 2011, 2014). However, there is a built-in assumption that the resolution is isotropic in nearest neighbour distances calculation, which is erroneous. This is what motivated Li and co-workers to exploit the higher resolution achieved at selected poles to quantify short-range ordering (Li et al., 2024, 2023), which appears as a more reasonable approach considering the possible extend of the aberrations and resulting intermixing in such alloy systems.

For studying small clusters and precipitates, there have been many debates as to the optimal parameters to use for the search via a range of algorithms (Marquis & Hyde, 2010; Dumitraschkewitz et al., 2018), including via specific heuristics (Marceau et al., 2011), and through interlaboratory, round-robin kind of experiments (Dong et al., 2019). The difficulty in defining these parameters arises from the fact that atoms within a cluster, i.e. that are in the near-neighbourhood of each other within the material, will not necessarily remain so in the reconstructed data due to the aberrations in their trajectories. This makes the distinction of clusters from the matrix as well as the measurement of their composition challenging as the ions from the matrix and clusters overlap, thereby blurring the compositional fields (Blavette et al., 2001). Recent efforts to quantify the influence resolution on the composition of precipitates by Jenkins et al. are noteworthy (Jenkins et al., 2024), yet their measurements of the resolution within the matrix may not readily apply to precipitates, since the aberrations affecting ion trajectories are composition-dependent (Vurpillot, Bostel, & Blavette, 2000; De

Geuser & Gault, 2020). This does not mean that the atoms of these particles or the particles themselves are not detected, simply that trajectory aberrations will make them appear typically larger, and their measured composition will be modified by overlaps with the matrix. This is a known issue, and there are possible ways to correct the measured composition to account for such effects (Blavette et al., 2001).

# Can we ever get an aberration-corrected atom probe?

Conversely to most conventional microscopes, aberrations in APT arise from the specimen itself, first at the atomic scale and then on the mesoscale, i.e. because of the shape of the specimen over distances that can range from nanometres almost to tens of microns. Both aspects are complex, with multiple origins, sometimes affecting atoms and ions over multiple scales.

Once again, these surface processes were suspected or known since over 50 years (Krishnaswamy et al., 1975; Waugh, A R et al., 1976; Krishnaswamy et al., 1977). Figure 9 showcases the difference in the achievable accuracy of imaging atomic positions by FIM and and through field evaporation as would be for APT. This well-known image obtained for pure Ir highlights the influence of the trajectory aberrations, the origins of which have been better understood from the extensive atomistic simulation work undertaken over the past two decades, by using finite-element methods (Vurpillot & Oberdorfer, 2015), boundary element methods and alternative mesoscale simulations (Loi et al., 2013; Rousseau et al., 2020; Fletcher et al., 2019; Hatzoglou & Vurpillot, 2019) or density-functional theory (Sanchez et al., 2004; Yao et al., 2015; Ashton et al., 2020; Katnagallu et al., 2023; Katnagallu, 2025). To get a holistic view on these aberrations will require integrating the modelling from the atomic- to the mesoscale. These models will also need to be confronted with experiments, in order to be validated. Yet recent evidence (Shaikh et al., 2025) suggests that stochastic fluctuations in the lateral velocity along with the roll-up of atoms on their neighbours at the surface may preclude from ever reaching full atomic-resolution, even in the case of a pure metal.

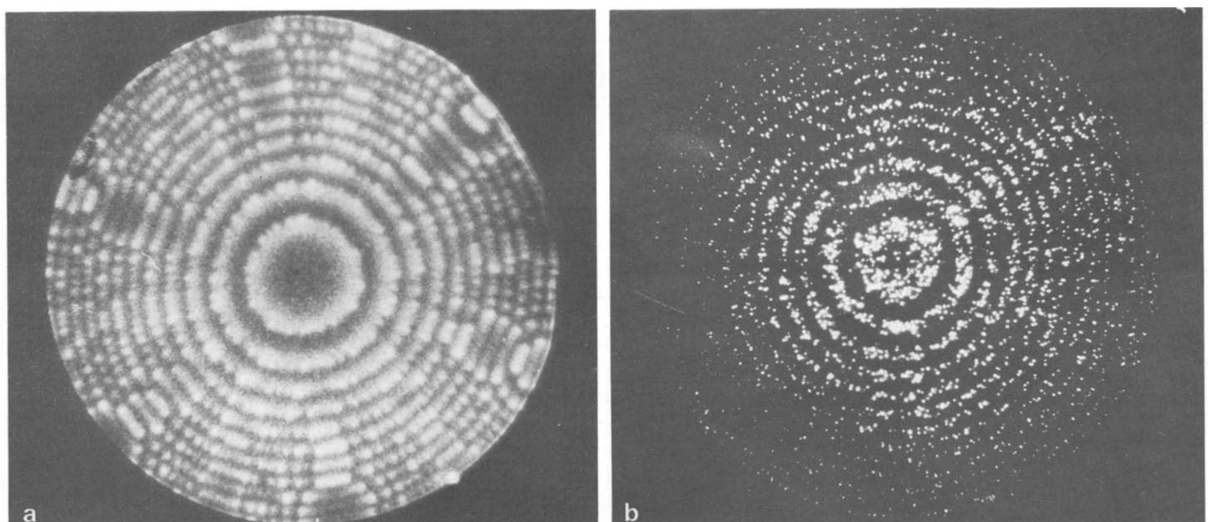

*Figure 9:* (a) Helium-ion image of an iridium specimen at 78 K. (b) Field desorption by pulsed-field-evaporation of a fraction of a monolayer image formed of iridium from the surface (a).

On the mesoscale, there is hope that actual quantification of the electric field distribution near APT specimens across scales can be achieved (Beleggia et al., 2014). This could be facilitated by the development of integrated instruments combining APT and transmission-electron microscopy (Mayer et al., 2023; Da Costa et al., 2024). These measurements will likely remain challenging and may not readily provide the resolution up to the atomic scale that is needed, requiring targeted novel research. Finally, the grounds for integrating such multiscale information to improve the reconstruction protocols have now been established in principle (N Rolland et al., 2015; Fletcher et al., 2020, 2022; Hatzoglou et al., 2025, 2023).

The caveat remains that the roll-up motion of the departing, partly charged atoms onto its neighbour is likely generally occurs and is stochastic in nature – this implies that there are multiple possible paths that are energetically accessible to each departing atom, hindering the repositioning of each atom back onto its exact location on the surface (Ashton et al., 2020). This roll-up can be extreme in some cases – see the recent work by Katnagallu et al. on Li for instance (Katnagallu et al., 2025). A most likely position may be determined, and for such problems, maybe deploying artificial intelligence methods will be necessary, yet this will lead to a reconstruction of the most probable position of an atom, not unlikely what is typically done in single-particle cryo-transmission electron microscopy using e.g. Alphafold (Abramson et al., 2024). This would remove the impression that APT is deterministic, and may be a positive development for the community to consider atoms within the reconstruction as a density of probability of their location.

## Summary & conclusions

We have here provided a critical perspective on the state-of-the-art of the knowledge on the spatial resolution in APT. Once again, this is not meant to be critical to the technique in general, simply to map the metrology of the technique and its capabilities. Table 1 is an attempt at summarising the data discussed herein. In the end, in a general case, one cannot say more than "the resolution is typically better than 1 nm in 3D". And even this may sometimes not be completely correct. And once again, this is fine, as achieving such a spatial resolution in 3D with a chemical sensitivity in the range of tens of ppm remains impossible for almost any other technique. Yet, we must accept, as a community that quoting a single value is misleading and should be avoided.

| Cases | Sample | Running conditions | Microscope geometry | $\Delta_{eff}$ | Occurrence |
|---|---|---|---|---|---|
| **Optimal** | Pure metals small specimen | Voltage pulsing, low specimen temperature | Straight flight path | 0.2 – 0.3 nm | Very uncommon |
| **Favourable** | Pure and low alloyed metals Small specimen | Voltage or laser pulsing (low laser pulse energy), low specimen temperature | Straight flight path | 0.4 – 0.7 nm | Uncommon |
| **General** | Metals and doped single phase semiconductors | Voltage or laser pulsing with low laser pulse energy, low specimen temperature | Straight flight path or with mesh (lens / reflectron) | ca. 1 nm | Most common |
| **Non-favourable cases** | Metals & non-metals in general, complex semiconductors, oxides | Laser up to high laser pulse energy or voltage pulsing at high base temperature | Straight flight path or with mesh (lens / reflectron) | > 1 – 2 nm | Common |

*Table 1: summary table of the achievable effective resolution for different sample types, running conditions and microscope geometries.*

## Conflicts of Interest

The authors declare no conflicts of interest.